# The phase composition and physical properties of melt-quenched multicomponent alloy $FeCoNiB_{0.7}Si_{0.3}Be$

O.I. Kushnerov[1], S.I. Ryabtsev[1], P.O. Galagan[1], V.F. Bashev[2]

[1]Oles Honchar Dnipro National University, Dnipro, Ukraine

[2]Dniprovsky State Technical University, Kamianske, Ukraine

The structure and physical properties of the high-entropy multicomponent alloy $FeCoNiB_{0.7}Si_{0.3}Be$ in the as-cast and melt-quenched states were studied. The cooling rate of the melt-quenched films was estimated to be ~ $10^6$ K/s based on the film thickness. X-ray analysis revealed a multiphase structure, including a BCC-type ordered phase (structural type B2) and intermetallic compounds $(Fe, Ni, Co)_2B$. In melt-quenched samples, the fraction of the B2 phase increased, leading to a decrease in microhardness (from 10400 MPa in as-cast to 8900 MPa in melt-quenched samples). Magnetic studies confirmed the ferromagnetic nature of the $FeCoNiB_{0.7}Si_{0.3}Be$ alloy. The coercive field of melt-quenched samples (17500 A/m) was significantly higher than that of the as-cast ones (5200 A/m), which is attributed to structure refinement and the increase in the level of microstresses.

Keywords: high-entropy alloy, structure, phase composition, melt-quenching, mechanical properties, magnetic properties.

## 1. *Introduction*

The traditional approach to alloy development involves selecting a base material based on the primary requirements of the final product and adding alloying elements to achieve the desired technological and operational properties. This method has led to the creation of many multicomponent alloys based on a single component and, in some cases, on two or three components. Such alloys have been extensively studied and optimized to meet specific mechanical, thermal, and chemical requirements. However, over the past two decades, this approach has undergone fundamental changes due to the advent of new alloying strategies aimed at expanding the design space for novel materials with enhanced properties.

A new alloy design principle has emerged, based on combining several principal elements (typically five or more) in relatively high concentrations (5 – 35 at. %), often in equimolar proportions [1-3]. This concept has led to the development of a new class of materials known as high-entropy alloys (HEAs). It was initially hypothesized that the high configurational entropy of mixing in these alloys would promote the formation of simple crystalline phases rather than complex intermetallic compounds. This idea is rooted in thermodynamic principles, where a higher entropy state favours phase stability, thereby reducing the tendency of these alloys to form brittle intermetallic phases. Subsequent studies have expanded this initial concept, revealing the potential for obtaining extremely diverse HEA structures, ranging from simple and ordered solid solutions to amorphous states. The phase composition of these alloys may include not only body-centred cubic (BCC) or face-centred cubic (FCC) solid solutions but also intermetallic compounds, depending on the atomic size difference, valence electron concentration, and mixing enthalpy of the constituent elements. This structural diversity enables the development of unique materials that integrate phases with different physical properties, making them highly promising for various engineering applications, including aerospace, biomedical, and energy storage technologies [1-8].

Among the strategies for improving HEA properties, rapid solidification (cooling rates exceeding $10^4$ K/s) has attracted particular attention. This method, known as melt quenching, enables the formation of non-equilibrium structures with unique characteristics, such as extended solid solubility, suppressed phase segregation, and refined microstructures. The influence of cooling rate on phase stability is a critical factor in determining the final properties of HEAs, as it directly affects grain size, defect concentration, and phase evolution [9-11]. Moreover, rapid cooling can induce glass-forming tendencies in certain HEA compositions, leading to the formation of bulk metallic glasses (BMGs) with superior strength, hardness, and wear resistance. These amorphous HEAs have garnered significant interest for their potential use in structural and functional applications where high strength-to-weight ratio and corrosion resistance are required.

Particular attention has been drawn to the potential of HEAs as functional materials, especially in the field of magnetic materials [9-14]. Many HEAs exhibit tuneable magnetic behaviour due to the presence of transition metals such as Fe, Co, and Ni, which contribute to ferromagnetism, while non-magnetic elements influence the Curie temperature, saturation magnetization, and coercivity. Such properties make HEAs suitable for applications in electromagnetic shielding, soft magnetic materials, and magnetocaloric refrigeration [12-14].

In this study, we investigate the influence of cooling rate on the characteristics of the $FeCoNiB_{0.7}Si_{0.3}Be$ multicomponent HEA. High-entropy alloys of the Fe-Co-Ni-B-Si system have already been described in the literature, particularly $FeCoNiB_{0.7}Si_{0.3}$ is a high-entropy metallic glass [15-17]. In this work, we added beryllium to this system, due to its advantageous physical characteristics such as high hardness and low density, as well as high glass-forming ability. The addition of Be is expected to influence the thermal stability, mechanical strength, and glass-forming ability of the alloy. Understanding the impact of rapid cooling on the structural and functional properties of this alloy is essential for further optimizing its properties for advanced engineering applications.

## 2. *Experimental*

As-cast HEA samples of $FeCoNiB_{0.7}Si_{0.3}Be$ with a nominal composition of 20 at.% Fe, 20 at.% Co, 20 at.% Ni, 14 at.% B, 6 at.% Si and 20 at.% Be were obtained using a high-temperature Tamman furnace in an argon flow. The average cooling rate in the copper mold was approximately ~$10^2$ K/s. To ensure homogeneity, the obtained ingots were remelted at least three times. Some of the as-cast samples were then used to produce melt-quenched (MQ) films using a splat- quenching technique. A technique for splat quenching used in the present work consisted of the rapid cooling of melt drops upon their collision with the internal surface of a rapidly rotating (~8000 RPM) hollow cylinder of copper. The linear velocity of the inner surface of the cylinder reached 110 m/s. The cooling rate $V_c$ was estimated using the equation [10]

$$V_c = \frac{\alpha}{c\rho\delta}(T - T_0), \quad (1)$$

where $c$ is the heat capacity, ρ is the density, α is the thermal conductivity coefficient, $T$ is the film temperature, $T_0$ is the ambient temperature, and δ is the film thickness. Given the obtained film thickness of ~40 μm, the calculated cooling rate was approximately ~$10^6$ K/s.

X-ray diffraction (XRD) analysis of both as-cast samples and melt-quenched films was performed using a DRON-2.0 diffractometer with monochromatic Cu Kα radiation. The diffraction patterns were analysed using QualX2 software for phase identification [18].

The magnetic properties of the samples were measured using a vibrating sample magnetometer (VSM-1) at room temperature. The microhardness was examined using a PMT-3 tester under a load of 100 g.

## 3. *Results and Discussions*

During the study of HEAs, many parameters have been proposed that can be used to describe their thermodynamic, electronic, and structural characteristics [1-3]. These include configurational entropy of mixing $\Delta S_{mix}$, enthalpy of mixing $\Delta H_{mix}$, thermodynamic parameter Ω, topological parameter δ$r$ which represents the difference in atomic radii of the alloy components, and valence electron concentration (VEC) which accounts for the concentration of valence (s+d orbital) electrons per atom.

$$\Delta S_{mix} = -R \sum_{i=1}^{n} c_i \ln c_i, \quad (2)$$

$c_i$- atomic fraction of the $i$-th component, $R$ - universal gas constant, $n$ is the number of alloy components.

$$\Delta H_{mix} = \sum_{i,j=1,i\neq j}^{n} \omega_{ij} c_i c_j, \quad (3)$$

where the regular melt-interaction parameter between $i$-th and $j$-th elements $\omega_{ij} = 4\Delta H_{mix}^{AB}$, and $\Delta H_{mix}^{AB}$- mixing enthalpy of binary liquid AB alloy.

$$\Omega = \frac{T_m \Delta S_{mix}}{|\Delta H_{mix}|} \quad (4)$$

where $T_m$ is the average melting temperature of the alloy $T_m = \sum_{i=1}^{n} c_i (T_m)_i$.

$$\delta r = \sqrt{\sum_{i=1}^{n} c_i \left(1 - \frac{r_i}{\bar{r}}\right)^2}, \quad (5)$$

where $\bar{r} = \sum_{i=1}^{n} c_i r_i$; $r_i$ - the atomic radius of the $i$-th element.

$$VEC = \sum_{i=1}^{n} c_i (VEC)_i, \quad (6)$$

Additionally, the difference in Allen electronegativities $\delta\chi^A$ was introduced to separate solid solution structures from intermetallic ones [19].

$$\delta\chi^A = \sqrt{\sum_{i=1}^{n} c_i (1 - \frac{\chi_i^A}{\chi^A})^2}, \quad (7)$$

where $\chi_i^A$ – Allen electronegativity of the $i$-th component.

Using data from [3,20,21] we calculated the values of these parameters (Tab.1).

Tab. 1. Electronic, thermodynamic, and structural parameters of the $FeCoNiB_{0.7}Si_{0.3}Be$ high-entropy alloy.

| $\Delta S_{mix}$, J/(mol·K) | $\Delta H_{mix}$, kJ/mol | Ω | VEC | $\delta r$, % | $\delta\chi^A$, % |
|---|---|---|---|---|---|
| 14.4 | -17.3 | 1.5 | 6.46 | 12.47 | 7.838 |

Based on the calculated values of these parameters and considering the known criteria from the literature [1-3, 22-24] for predicting the phase composition of HEAs, it can be concluded that the $FeCoNiB_{0.7}Si_{0.3}Be$ alloy should contain amorphous or intermetallic phases as well as a BCC-type solid solution. However, it should be noted that HEA design rules based on statistical results or average properties, or their deviations, are necessary but not sufficient conditions, especially at high cooling rates. In this case, the kinetic effect should be considered [25].

The analysis of diffraction patterns for as-cast and MQ samples of the $FeCoNiB_{0.7}Si_{0.3}Be$ alloy (Fig. 1) revealed a multiphase structure.

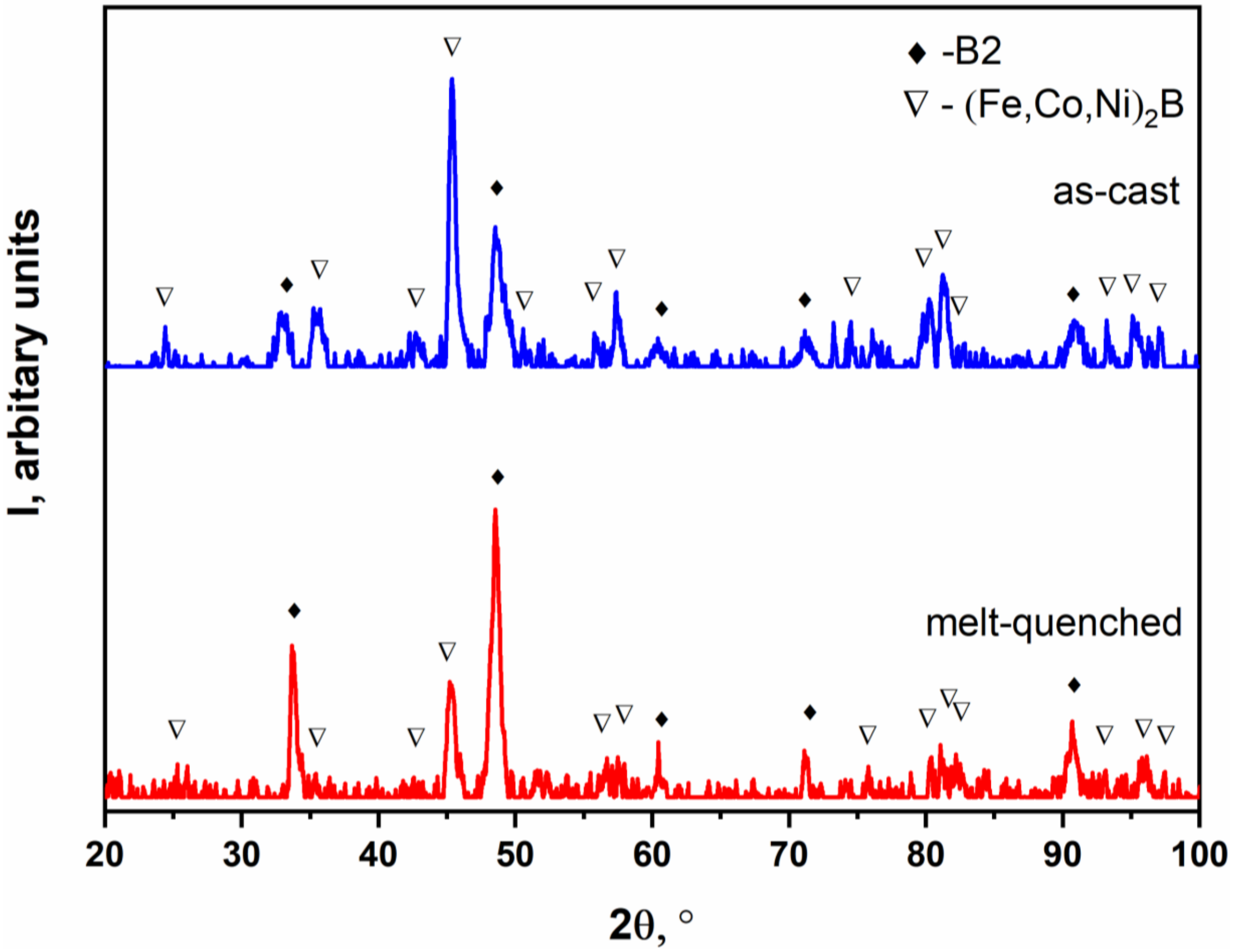

Fig.1. XRD patterns of as-cast and MQ samples of the $FeCoNiB_{0.7}Si_{0.3}Be$ HEA.

The as-cast sample exhibited peaks corresponding to an ordered solid solution with a BCC lattice (structural type B2, lattice parameter *a* = 0.2655 nm). Additionally, the intermetallic phase $(Fe, Ni, Co)_2B$ with a tetragonal lattice was present. In the MQ sample, the same phases were observed, but the lattice parameter of the ordered solid solution with a BCC lattice was *a* = 0.2653 nm. The intensity of the corresponding peaks indicated that the MQ sample had a higher content of the B2 phase compared to the $(Fe, Ni, Co)_2B$ phase. This confirms the fact, that rapid cooling of the melt can prevent the formation of complex intermetallics in HEAs.

The microstructures of the alloy samples were studied using optical and electron microscopes. To reveal the microstructure, the sample surfaces were treated with aqua regia. The obtained microstructure images are shown in Fig. 2. Analysis of these images indicates that the as-cast alloy is characterized by a dendritic structure with spinodal decomposition. The size of the dendrites is determined by the cooling rate of the melt. In MQ samples, a finely dispersed conglomerate of phases is formed due to the high cooling rate.

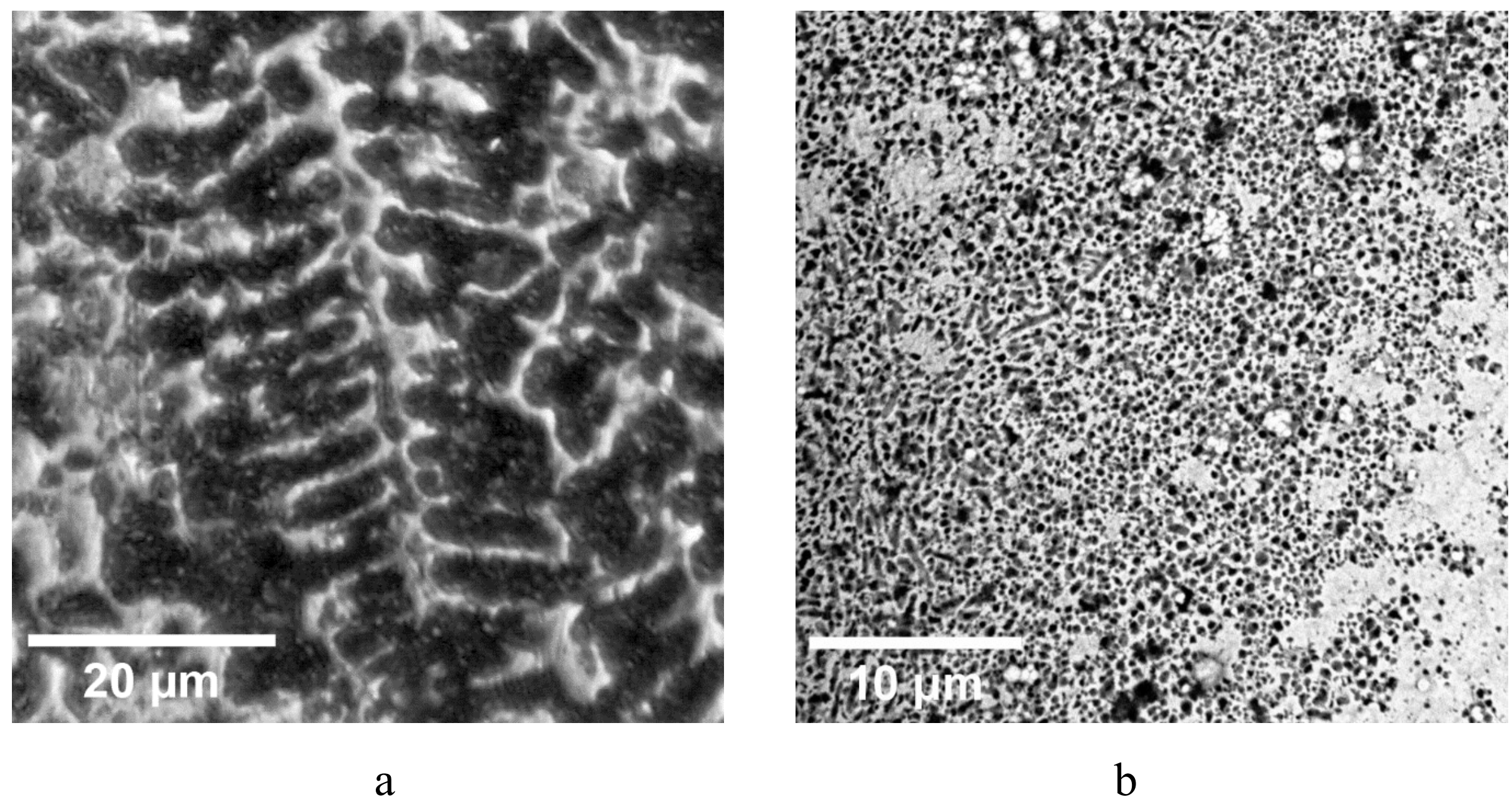

Fig. 2. Optical micrograph of the as-cast (a) and SEM image of the melt-quenched (b) $FeCoNiB_{0.7}Si_{0.3}Be$ HEA samples.

Comparative analysis of the microhardness of the $FeCoNiB_{0.7}Si_{0.3}Be$ HEA in different states revealed significant differences between the samples. The as-cast samples exhibited a microhardness of $H\mu$ = 10400±300 MPa, which exceeds the values of the MQ samples (8900±300 MPa). The high microhardness of the studied alloy is evidently due to the presence of an ordered solid solution B2 and intermetallic phases in its structure, which are characterized by very high hardness values.

The investigation of the magnetic properties of the $FeCoNiB_{0.7}Si_{0.3}Be$ alloy confirmed its ferromagnetic nature. Fig. 3 shows the room temperature hysteresis loops of the as-cast and MQ $FeCoNiB_{0.7}Si_{0.3}Be$ HEA.

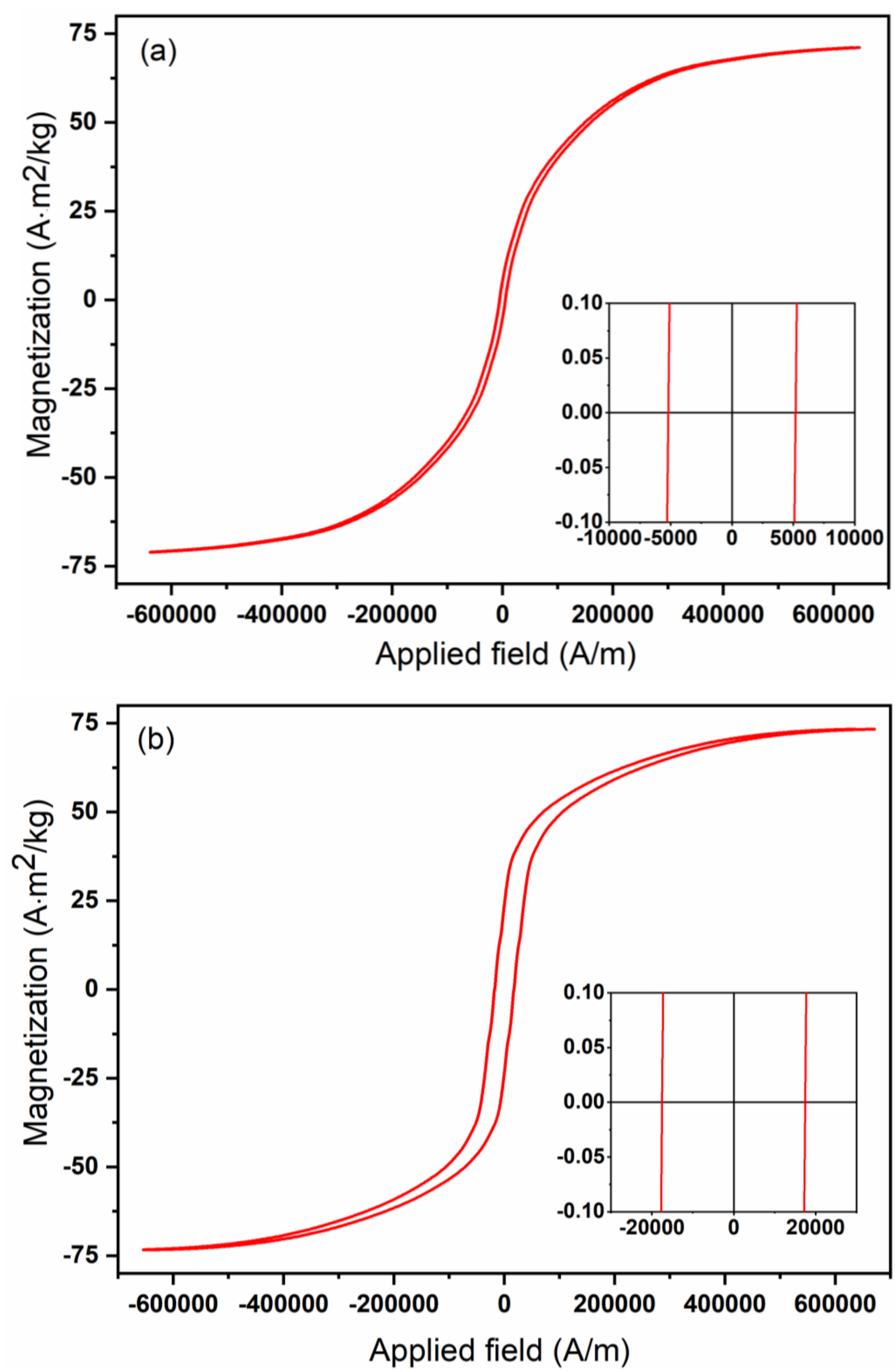

Fig.3. Hysteresis loops of as-cast (a) and melt-quenched (b) samples of $FeCoNiB_{0.7}Si_{0.3}Be$ HEA.

Based on the coercive field $H_c$, both types of samples can be classified as hard magnetic materials, with MQ samples demonstrating significantly higher $H_c$ values (17500 A/m) compared to as-cast samples (5200 A/m). At the same time, specific saturation magnetization $M_S$ remains practically unchanged with varying cooling rates, being 71 $A \cdot m^2/kg$ for as-cast and 73 $A \cdot m^2/kg$ for MQ samples. It can be explained by the fact that magnetization is primarily determined by the phase composition and type of crystal structure, which change slightly during quenching. The high coercive field values in MQ samples can be attributed to the refinement of the alloy structure with increased cooling rates. Additionally, domain wall movement is hindered due to the increased level of microstresses in the quenched films, as well as the presence of defects and nanoscale inclusions.

## 4. *Conclusions*

For the first time, a multicomponent high-entropy alloy $FeCoNiB_{0.7}Si_{0.3}Be$ with a multiphase structure has been obtained, containing an ordered BCC solid solution (B2 type) and an intermetallic phase $(Fe, Ni, Co)_2B$. Rapid quenching (up to $10^6$ K/s) increases the fraction of the B2 phase and reduces the content of the intermetallic phase.

The studied alloy, both in the cast and MQ states, exhibits well-defined ferromagnetic properties. Based on the coercive field magnitude, the $FeCoNiB_{0.7}Si_{0.3}Be$ HEA can be classified as a hard magnetic material.

The coercive field magnitude of the studied alloy increases with the cooling rate of the melt, which can be explained by the formation of a finely dispersed structure in the MQ films, as well as the presence of defects and nanoscale inclusions that hinder domain wall movement.

It has been confirmed that rapid quenching suppresses but does not completely prevent the formation of complex intermetallics in high-entropy alloys of the Fe-Co-Ni-B-Si-Be system.

Authors

Corresponding author

O.I. Kushnerov, Oles Honchar Dnipro National University

ORCID 0000-0002-9683-2041

e-mail: kushnrv@gmail.com

S. I. Ryabtsev, Oles Honchar Dnipro National University

ORCID 0000-0002-2889-5278

e-mail: siryabts@gmail.com

P. O. Galagan, Oles Honchar Dnipro National University

ORCID 0009-0007-5063-0616

e-mail: galaganpo@gmail.com

V. F. Bashev, Dniprovsky State Technical University

ORCID 0000-0002-3177-0935

e-mail: bashev_vf@ukr.net